\documentclass[11pt,a4paper]{article}
\usepackage{amsmath,amsfonts,amssymb,graphicx,color}

\numberwithin{equation}{section}

\renewcommand{\title}[1]{{\Large\bf\mbox{}\\\medskip#1\bigskip\medskip\\}}
\renewcommand{\author}[1]{{\large #1\smallskip\\}}
\newcommand{\address}[1]{{\em #1\medskip\\}}

\newcommand{\re}{\mbox{Re}}

\newcommand{\ir}{^{\mbox{\scriptsize IR}}}
\newcommand{\uv}{^{\mbox{\scriptsize UV}}}
\newcommand{\irt}{^{\mbox{\tiny IR}}}
\newcommand{\uvt}{^{\mbox{\tiny UV}}}
\newcommand{\sopra}[1]{\stackrel{\mbox{\footnotesize #1}}{\mapsto}}
\newcommand{\xilatt}{\xi_{\mbox{\scriptsize latt}}}
\newcommand{\gauss}[2]{\genfrac{[}{]}{0pt}{0}{#1}{#2}}

\makeatletter
\renewcommand{\@makecaption}[2]{
   \vskip\abovecaptionskip
   \sbox\@tempboxa{#1. #2}%
   \ifdim \wd\@tempboxa >\hsize
     #1. #2\par
   \else
     \global \@minipagefalse
     \hb@xt@\hsize{\hfil\box\@tempboxa\hfil}%
   \fi
   \vskip\belowcaptionskip}
\makeatother

\begin{document}

\vspace*{-15mm}
\begin{flushright} SISSA 109/2003/FM
\end{flushright}
\begin{center}

\title{Exact $(d) \mapsto (+)\&(-)$  boundary flow 
in the tricritical Ising model}

\author{Giovanni Feverati\footnote{feverati@sissa.it}}

\address{International School for Advanced Studies, Trieste, Italy}

\end{center}

\setcounter{footnote}{0}

\begin{abstract}
The integrable perturbation of the degenerate boundary condition $(d)$ by 
the $\varphi_{1,3}$ boundary field generates a renormalization group 
flow down to the superposition of Cardy boundary states $(+)\&(-)$.
Exact Thermodynamic Bethe Ansatz (TBA) equations for all the excited states
are derived here extending the results of \cite{FPR2} to this case.  
As an intermediate step, the non-Cardy boundary conformal sector 
$(+)\&(-)$ is also described as the scaling limit of an $A_4$ lattice model 
with appropriate integrable boundary conditions and produces  
the first example of superposition of
finitized Virasoro characters.
\end{abstract}
\section{Introduction}
Two dimensional quantum field theories in presence of a boundary have received 
an increasing attention in the last few years because of their 
important role in condensed matter physics (especially on quantum 
impurity problems) and in string theory (D-branes). 
Their integrable or conformal properties lead to the computation
of a number of exact results (boundary S-matrices \cite{gz}, 
conformal boundary states \cite{Cardy}, etc. See also \cite{GRW} and
references indicated there). 
An interesting problem is to connect different boundary conformal 
field theories (BCFT) by integrable renormalization group flows generated by
boundary operators.

The tricritical Ising model is the conformal unitary minimal model with
central charge $c=7/10$. Its phase diagram in presence of boundaries, with 
the bulk of the system maintained at the critical point, contains various 
fixed points (that are precisely boundary conformal field theories) connected 
by renormalization group flows \cite{Chim, affleck}.
Among them, those generated by perturbing a fixed point with the 
magnetic boundary field $\varphi_{1,3}$ are integrable. One of the two 
flows departing from the BCFT $(d)$ moves to a superposition of 
Cardy-type boundary states, $(+)\&(-)$. This is the simplest occurrence of 
the general observation \cite{RRS} that the boundary interactions do not 
make a distinction between pure states or superposition of states.
Actually, increasing the central charge of the minimal models, one expect 
more and more the appearance of superposition of states \cite{fs} so that 
the analysis presented here is a crucial step toward a quite general
phenomenon. 
Physically speaking, boundary correlation functions of pure states satisfy 
the cluster property while this is not true for superposition of states.
This is an indication of a first order phase transition occurring close to 
a BCFT that is not a pure state \cite{affleck}. 

The flow  $(d)=(2,2) \mapsto (+)\&(-)=(1,1)\oplus(3,1)$ is studied here 
using a lattice approach to formulate TBA equations, as described with 
great detail in \cite{FPR2} on which the present article is heavily built. 
For convenience, the same notations will be used.
This flow was identified in \cite{Chim} by the boundary S matrix; it 
preserves supersymmetry \cite{nep}; a TBA system for the ground state only 
was given \cite{NA} in a different channel (periodic).

The plan of the paper is as follows. 
In Section~\ref{s_critpnt}, the pattern of zeros for the boundary 
critical point \mbox{$(1,1)\oplus(3,1)$} on a finite lattice 
will be described and the finitized partition function will appear
as a superposition of finitized characters \cite{FinChar}.
In Section~\ref{s_3mech}, the flow will be recognized to be in 
the ``variable $r$'' family, and the mapping between finitized characters 
will be provided.
In Section~\ref{s_TBA}, the specific set of TBA equations for 
this flow will be derived and numerically solved.
In Appendix~A there is a summary of properties of Gaussian polynomials.
In Appendix~B the expression for the energy at the BCFT
$(1,1)\oplus(3,1)$ will be proved.
In Appendix~C the selection rules for the integration constants will be given,
that also apply to all the cases discussed in \cite{FPR2}.

\section{The boundary critical point $(1,1)\oplus(3,1)$\label{s_critpnt}}
An $N$ faces double row transfer matrix 
with boundary conditions $(r_i,a_i),\quad i=1,2$ on the left/right boundaries
is indicated by 
$\mathbf{D}^{N}_{r_1,a_1|r_2,a_2}(u,\xi_1,\xi_2)$ where $0<u<\lambda$ is the 
spectral parameter, $\lambda=\frac{\pi}{5}$ is the crossing parameter
and $\xi_i$ are parameters associated to the boundary weights.
If $\xi$ is chosen appropriately, the lattice boundary 
weight $(r,a)$ scales to the conformal boundary condition \cite{OPW, BP}
also labeled by $(r,a)$ \cite{BPPZ}.
In the rest of the paper, the only case that will be used is:
\begin{equation} \label{drtm}
\displaystyle \mathbf{D}^{N}(u,\xi_1)=
\mathbf{D}^{N}_{2,1|2,1}(u,\xi_1,\frac{3}{2}\lambda).
\end{equation}
The parameter $\xi_2$ will be kept fixed at the indicated value 
corresponding to the conformal boundary condition $(2,1)$.

From \cite{BP}, it is known that the continuum scaling limit of the 
double row transfer matrix 
$\mathbf{D}^{N}(\frac{\lambda}{2},\frac{3}{2}\lambda)$ is a realization 
of the $(2,1)\otimes(2,1)=(1,1)\oplus(3,1)$ conformal boundary 
conditions. This is obtained if the boundary parameter 
is chosen inside the real interval $\xi_1 \in [\lambda/2,5\lambda/2]$. 
The choice of the middle point of the interval will be useful later when 
moving off criticality by the introduction of an imaginary part for $\xi_1$.
The number of faces $N$ is required to be even by the adjacency conditions.

The pattern of the zeros of the eigenvalues $D(u)$ of 
$\mathbf{D}(u)$ is relevant to the classification of the states, 
both for the lattice and for the continuum theory, and must be determined with 
numerical observations performed on lattices of small 
size\footnote{The number of faces is 
mainly limited by the huge amount of RAM required to write the 
transfer matrix.} $N=12$ (in \cite{FPR2} the name ``numerics on D'' 
was used).
As typical for the $A_4$ lattice model \cite{OPW, PCA}, there are 
two analyticity strips 
$$
\mbox{strip 1: \quad} -\frac{\lambda}{2}<\re(u)<\frac{3\lambda}{2},
\qquad \qquad
\mbox{strip 2: \quad} 2\lambda<\re(u)<4 \lambda
$$ 
and two types of strings.
The 1-strings are single zeros
appearing in the center of each strip
$$ 
\re(u)=\frac{\lambda}{2} \mbox{~ or ~} 3\lambda
$$ 
while the 2-strings are pairs of zeros $(u,u')$ appearing on the edges of a
strip, with the same imaginary part
$$ 
(\re(u),\re(u'))= \left\{
\begin{array}{lc} (-\lambda/2,3\lambda/2), ~~~& \mbox{strip 1,}\\[2mm]
(2\lambda,4\lambda), & \mbox{strip 2.}
\end{array} \right.
$$ 
In the upper half plane, let $m_1$, $m_2$ be the number of 1-strings 
and $n_1$, $n_2$ the number of 2-strings in each strip.
The following relations hold between the numbers of zeros:
\begin{equation}\label{n1n2} \left.
\begin{array}{ccl}
n_1 &=& \frac{N+m_2}{2}-m_1 \geqslant 0 \\[2mm]
n_2 &=& \frac{m_1}{2}-m_2+1 \geqslant 0
\end{array} \right\} \Rightarrow m_1,m_2 ~~\mbox{even.}
\end{equation}
The states on the lattice and at the scaling limit are uniquely characterized
by the {\em non-negative quantum numbers} $\{I_k^{(j)}\}$; they have
the topological meaning that for a given 1-string, 
$I_k^{(j)}$ is the number of 2-strings with larger coordinate. This remains 
true at the scaling limit. 

The determination of the scaling energy corresponding to each eigenvalue
is the crucial element to distinguish states in the $(1,1)$ or $(3,1)$ sectors.
Remembering that $\Delta_{1,1}=0,\quad \Delta_{3,1}=\frac{3}{2}$, the 
separation between levels is quite sharp. In spite of this,  
the finite size effects for $N=12$ faces are still dominant on the accuracy
of the energies, apart for the first few levels $E\leqslant 5$, so 
the complete distinction of states among the two sectors requires 
informations from the TBA equations that 
will be introduced later. 
Anticipating here these results, one has: 
\begin{eqnarray}
(1,1): & &I^{(2)}_1 < n_2 \quad \mbox{or} \quad m_2=0 ,\label{dist11} \\ 
(3,1): & &I^{(2)}_1 = n_2. \label{dist31}
\end{eqnarray}
This means that the discrimination is given by the lower object in the second 
strip: if it is a 2-string, the sector $(1,1)$ is selected, if it is
a 1-string the sector $(3,1)$ is selected.
The given conditions imply two useful necessary conditions for a state to be 
in the corresponding sector:
\begin{eqnarray} 
(1,1) && 0\leqslant I^{(2)}_1 < n_2\Rightarrow \frac{m_1}{2}-m_2\geqslant 0,
\label{nec11}\\
(3,1) & & m_2\geqslant 2 \Rightarrow m_1 \geqslant 2. \label{nec31}
\end{eqnarray}
The former is an enforcement of the second expression in (\ref{n1n2}).
The energy corresponding to each state is finally given by
\begin{eqnarray} \label{critenergy}
E&=&-\frac{c}{24}+\frac{\boldsymbol{m}^{T}C\boldsymbol{m}}{4}-\frac{1}{2} 
\delta_{I^{(2)}_1,n_2}+\sum_{i=1,2}\sum_{k=1}^{m_i} I_k^{(i)}, \\
C&=&\begin{pmatrix} 2 &-1\\-1&2\end{pmatrix}, \qquad \delta_{a,b}: 
~~\mbox{Kronecker's delta}, \qquad \boldsymbol{m}=
\begin{pmatrix}m_1\\m_2\end{pmatrix}. \nonumber
\end{eqnarray}
$C$ is precisely the $A_2$ Cartan matrix. 
The complete and independent proof of this expression will be given in
Appendix B.
 
Here a crucial observation can be done that there is a violation of the 
{\em topological rule for the energy}. Indeed, when only one sector is 
involved, if the relative position of a 1-string and a neighboring 
underlying 2-string 
are exchanged the energy increases of 1 unit. The opposite exchange reduces 
the energy of 1 unit \cite{OPW}. Presently two sectors 
are considered and this rule is violated in precisely one case: if the 
lower objects in the second strip are a 1-string and an underlying 2-string 
(i.e. $I_1^{(2)}=n_2-1$), their exchange increases the energy of $\frac{1}{2}$
units. The opposite exchange reduces the energy of $\frac{1}{2}$
units. This corresponds to the Kronecker's delta term in (\ref{critenergy}). In
all the other cases the usual rule holds.
Notice that this violation is expected from counting arguments, namely with 
the usual rule only, there is no way to match the states given by 
$\mathbf{D}(u)$ with the states contained in $(1,1)\oplus(3,1)$.
In Table~\ref{t22_1131} there is the first appearance of this violation, namely
the infrared state\footnote{The notation indicates the content of zeros: 
for example, $(3\,2\,2\,0|1\,0)$ is a state with 4 zeros in the first 
strip and 2 in the second; their quantum 
numbers are respectively $I_1^{m_1}=3, \quad I_2^{m_1}=2, \ldots$. 
If a frozen zero occurs \cite{OPW}, a parity $\sigma=\pm 1$ is added.} 
$(0\,0\,0\,0|0\,0)$, whose energy is $6$, has two excitations,
$(0\,0\,0\,0|1\,0)$, with energy $\frac{3}{2}+5$, and $(1\,0\,0\,0|0\,0)$,
with energy $7$.

The finitized partition function is defined as a sum on the states of a 
finite lattice where the energy of each state is fictitiously assumed to be
equal to its scaling limit energy:
\begin{equation}\label{finitZ}
Z_N(q)=q^{-\frac{c}{24}}\sum_{\mbox{\scriptsize states}}q^E 
=q^{-\frac{c}{24}}\sum_{m_1,m_2,\{I_k^{(i)}\}}
q^{\frac{1}{4}\boldsymbol{m}^{T}C\boldsymbol{m} -\frac{1}{2} 
\delta_{I^{(2)}_1,n_2}+\sum_{i,k}I_k^{(i)}} 
\end{equation} 
where $q=\exp(-2\pi M/N)$ is the modular parameter.
Now, it will be shown that the finitized partition function matches the sum of 
characters. 
The sum on states (\ref{finitZ}) can be split in two parts, for 
$I_1^{(2)}<n_2$ (sector $(1,1)$) and for $I_1^{(2)}=n_2$ (sector $(3,1)$). 
Using the Gaussian polynomials and their properties as in Appendix A leads to: 
\begin{eqnarray}
\mbox{if}~~ I_1^{(2)}<n_2 &\!:~&\sum_{0\leqslant I^{(2)}_{m_2}\leqslant 
\ldots \leqslant I_1^{(2)}<n_2} q^{I^{(2)}_1+\ldots I^{(2)}_{m_2}}=
\gauss{m_2+n_2-1}{m_2}, \\
\mbox{if}~~ I_1^{(2)}=n_2 &\!:~&\sum_{0\leqslant I^{(2)}_{m_2}\leqslant 
\ldots \leqslant I_1^{(2)}=n_2} q^{I^{(2)}_1+\ldots I^{(2)}_{m_2}}=q^{n_2}
\gauss{m_2+n_2-1}{m_2-1}.
\end{eqnarray}
The sum on the first strip is not modified by this counting. The partition 
function now reads
\begin{eqnarray} \label{partf}
Z_N(q) &=& q^{-\frac{c}{24}} \sum_{(1,1)} 
q^{\frac{1}{4}\boldsymbol{m}^{T}C\boldsymbol{m}}
\gauss{m_1+n_1}{m_1} \gauss{m_2+n_2-1}{m_2} \\
&&+\, q^{-\frac{c}{24}} \sum_{(3,1)} 
q^{\frac{1}{4}\boldsymbol{m}^{T}C\boldsymbol{m}-\frac{1}{2}}
q^{n_2}\gauss{m_1+n_1}{m_1} \gauss{m_2+n_2-1}{m_2-1}
\nonumber
\end{eqnarray}
where the labels on the sums indicate that the sums on $m_1,m_2$ are 
restricted by the constraints imposed by the corresponding sector 
(\ref{n1n2}-\ref{nec31}). 
The first sum is easily recognized to be the finitized character 
$\chi^{(N)}_{1,1}(q)$. In the second term, the redefinition of 
$m_2\rightarrow m_2'=m_2-1\geqslant 1$ and odd, leads to 
$$
\frac{1}{4}(m_1 ~~m_2)\,C\begin{pmatrix}m_1 \\m_2\end{pmatrix}+n_2-\frac{1}{2}=
\frac{1}{4}(m_1 ~~m_2')\,C\begin{pmatrix}m_1 \\m_2'\end{pmatrix}
$$
so that it becomes the finitized character 
$\chi^{(N)}_{3,1}(q)$.
This proves that the finitized partition function is the sum of the 
corresponding finitized characters:
\begin{equation}
Z_N(q)=\chi^{(N)}_{1,1}(q) +\chi^{(N)}_{3,1}(q)
\end{equation}
and, consequently, that the pattern of zeros (\ref{n1n2}) and the distinction 
of sectors (\ref{dist11}, \ref{dist31}) are consistent with the known 
scaling properties of the given transfer matrix.

\section{The flow: 3 mechanisms\label{s_3mech}}

\begin{figure}[ht]
\includegraphics[width=0.31\linewidth]{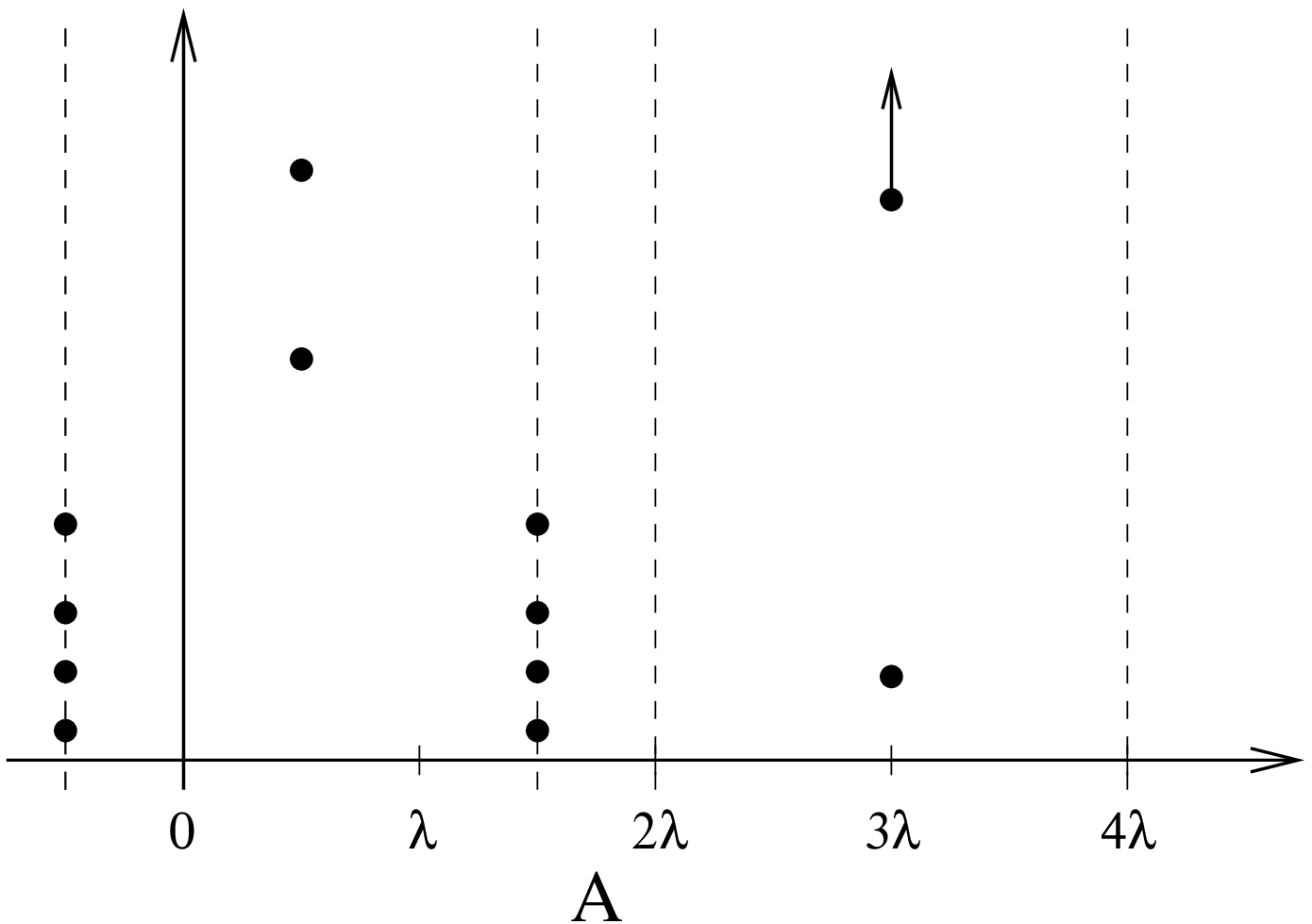}
\hfill\includegraphics[width=0.31\linewidth]{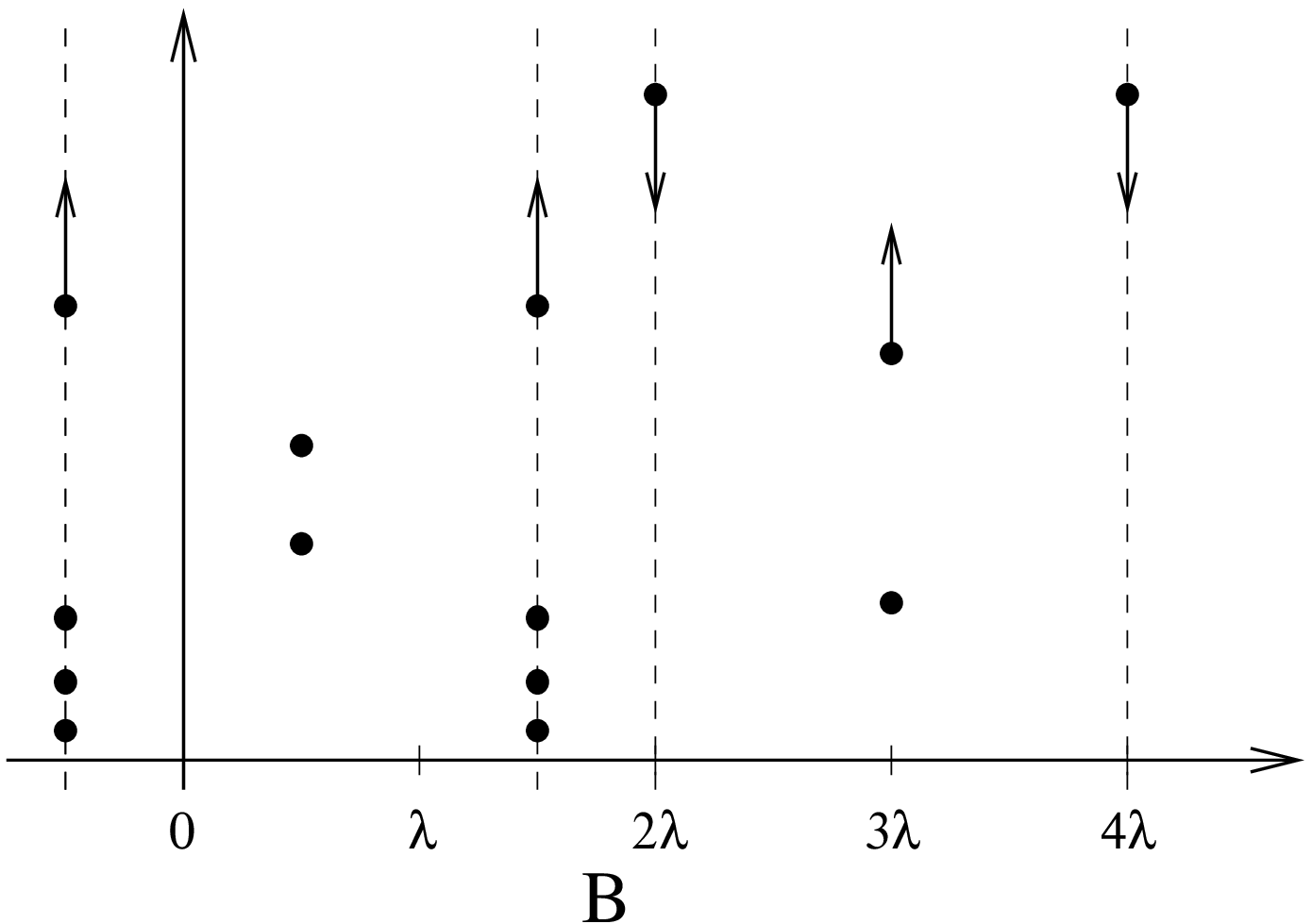}
\hfill\includegraphics[width=0.31\linewidth]{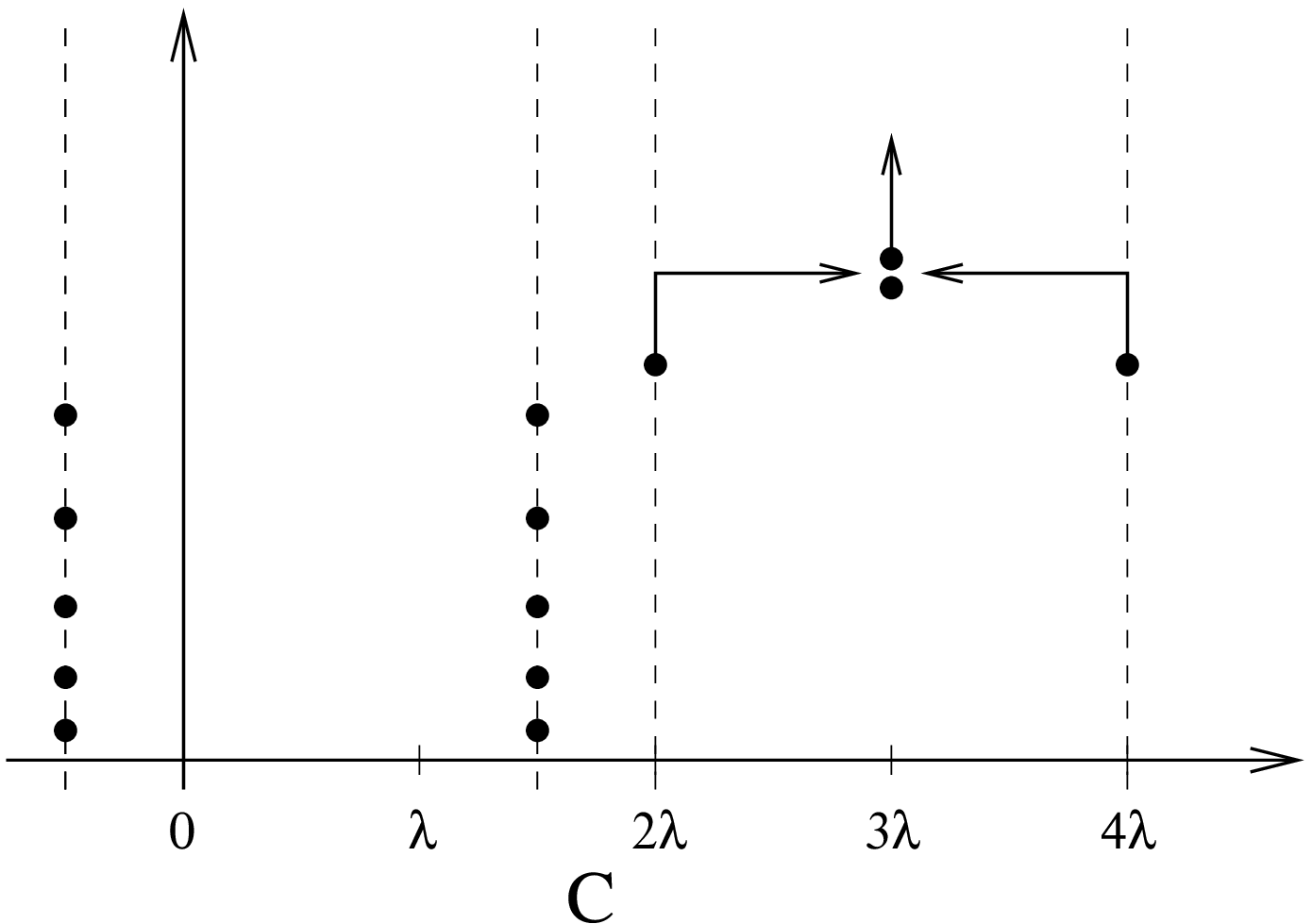}
\vspace*{-3mm}
\caption{\small \label{mechan} The three mechanisms A, B, C respectively
that change string content during the flow
\mbox{$\chi_{1,1}+\chi_{3,1}$} $\mapsto \chi_{2,2}$
which is the reverse of the physical flow.
These mechanisms are illustrated for the states:
A, $(0\,0|0\,0)\,\mapsto\,(0\,0|0)_{+}$;
B, $(1\,1|0\,0)\,\mapsto\,(0\,0|1)_{-}$;
C, $(~)\,\mapsto\,(~|0)_{-}$.}
\end{figure}

Considering the transfer matrix $\mathbf{D}(u,\xilatt)$ with 
$\xilatt=\frac{3}{2}\lambda+i\xi/5$, one can compute 
the expected endpoints, 
\begin{equation} \label{1321scaling}
\mathbf{D}^N (u,\frac{3}{2}\lambda+i\frac{\xi}{5})
\longrightarrow \left\{ \begin{array}{l@{\hspace{4mm}}l}
\chi_{2,2} & \mbox{if ~} \xi \rightarrow \pm \infty \\[3mm]
\chi_{1,1}+\chi_{3,1} & \mbox{if ~} \xi =0  \end{array}
\right.
\end{equation}
as it was done in \cite{FPR2} for various types of boundary.
The choice of the real part $\re(\xilatt)=\frac{3}{2}\lambda$ makes the
double row transfer matrix real symmetric on the center of each strip, 
$\re(u)=\frac{\lambda}{2}, ~3\lambda$ so that the eigenvalues and the energies
are real. Any different value for the real part would lead to complex energies
and non-unitary flows.
Summarizing, one has integrability, correct endpoints, unitarity, 
and uniqueness so the identification of this lattice flow with 
the renormalization flow $(2,2)\rightarrow(1,1)\oplus(3,1)$ is completely
natural.

From extensive ``numerics on $D$'' one can figure out the correspondence 
between infrared (IR) and ultraviolet (UV) states, Table~\ref{t22_1131}, 
and the three 
mechanisms A, B, C active to change the pattern of zeros.
They are the same known for the ``variable $r$'' flows in \cite{FPR2}
so the present flow is part of that family.
%
For their complete description the reader is referred to \cite{FPR2}  
while here there is a summary of the properties and in Fig.~\ref{mechan} 
there is an example of their action.

\begin{table}[t]
\caption{\label{t22_1131}
\small
Flow $\chi_{1,1}+\chi_{3,1} \mapsto \chi_{2,2}$ (reverse of the physical flow).
The explicit mapping of states from IR to UV up to the IR 
level 7 are presented here.
$n\ir,\:n\uv$ are the excitation levels
above the ground states, respectively $\Delta_{1,1}=0$ and 
$\Delta_{2,2}=\frac{3}{80}$. 
The IR states in the two sectors are easily recognized because in 
$(3,1)$ they are separated by $\Delta_{3,1}=\frac{3}{2}$ from the 
true ground state.}\vspace*{-7mm}
\begin{center}$$
\begin{array}{|c|rcl|c||c|rcl|c|}
\hline 
n\ir\rule[-4mm]{0mm}{10mm}& \multicolumn{3}{c|}{\mbox{\begin{tabular}{c}mapping
of states, \\ mechanism\end{tabular}}} & n\uv &
   n\ir & \multicolumn{3}{c|}{\mbox{\begin{tabular}{c}mapping of states, \\ 
mechanism\end{tabular}}} & n\uv\\
\hline
0\rule[-1mm]{0mm}{6mm} & (~) & \sopra{C} & (~|0)_{-} &0&
         \frac{3}{2}+4 & (3\,1|0\,0) &\sopra{B} & (2\,0|1)_{-} & 5 \\[1mm]
\frac{3}{2} & (0\,0|0\,0) & \sopra{A} & (0\,0|0)_{+} & 1 &
         \frac{3}{2}+4 & (2\,2|0\,0) &\sopra{B} & (1\,1|1)_{-} &  5 \\[1mm]
2 & (0\,0) & \sopra{C} & (0\,0|0)_{-} &  2 &
         6 & (0\,0\,0\,0|0\,0) &\sopra{A} & (0\,0\,0\,0|0)_{+} & 5 \\[1mm]
\frac{3}{2}+1 & (1\,0|0\,0) & \sopra{A} & (1\,0|0)_{+} & 2 &
         6 & (4\,0) &\sopra{C} & (4\,0|0)_{-} & 6 \\[1mm]
3 & (1\,0) &\sopra{C} & (1\,0|0)_{-} & 3 &
         6 & (3\,1) &\sopra{C} & (3\,1|0)_{-} & 6 \\[1mm]
\frac{3}{2}+2 & (2\,0|0\,0) &\sopra{A} & (2\,0|0)_{+} & 3 &
         6 & (2\,2) &\sopra{C} & (2\,2|0)_{-} & 6 \\[1mm]
\frac{3}{2}+2 & (1\,1|0\,0) &\sopra{B} & (0\,0|1)_{-} & 3 &
         \frac{3}{2}+5 &(0\,0\,0\,0|1\,0)&\sopra{A}&(0\,0\,0\,0|1)_{+}&6\\[1mm]
4 & (2\,0) &\sopra{C} & (2\,0|0)_{-} & 4 &
         \frac{3}{2}+5 & (5\,0|0\,0) &\sopra{A} &(5\,0|0)_{+} & 6 \\[1mm]
4 & (1\,1) &\sopra{C} & (1\,1|0)_{-} &  4 &
         \frac{3}{2}+5 & (4\,1|0\,0) &\sopra{B} & (3\,0|1)_{-} & 6 \\[1mm]
\frac{3}{2}+3 & (3\,0|0\,0) &\sopra{A} & (3\,0|0)_{+} & 4 &
         \frac{3}{2}+5 & (3\,2|0\,0) &\sopra{B} & (2\,1|1)_{-} & 6 \\[1mm]
\frac{3}{2}+3 & (2\,1|0\,0) &\sopra{B} & (1\,0|1)_{-} & 4 &
         7 & (1\,0\,0\,0|0\,0) &\sopra{A} & (1\,0\,0\,0|0)_{+} & 6 \\[1mm]
5 & (3\,0) &\sopra{C} & (3\,0|0)_{-} & 5 &
         7 & (5\,0) &\sopra{C} & (5\,0|0)_{-} & 7 \\[1mm]
5 & (2\,1) &\sopra{C} & (2\,1|0)_{-} & 5 &
         7 & (4\,1) &\sopra{C} & (4\,1|0)_{-} & 7 \\[1mm]
\frac{3}{2}+4 & (4\,0|0\,0) &\sopra{A} & (4\,0|0)_{+} & 5 &
         7 & (3\,2) &\sopra{C} & (3\,2|0)_{-}  & 7 \\[1mm]
\hline
\end{array}$$
\end{center}
\end{table}

The mechanisms A,B,C force the following changes for
the various parameters:
\begin{list}{}{\itemindent 12mm}
\item[A. $I^{(1)}_{m_1}={I^{(2)}_{m_2\irt}}\ir=0$:]
\begin{eqnarray} \label{mech2A}
m_2\ir &\mapsto& m_2\uv=m_2\ir-1, \qquad \sigma=1.
\end{eqnarray}
\item[B. ${I^{(1)}_{m_1}}\ir> 0, \quad {I^{(2)}_{m_2\irt}}\ir=0$:]
\begin{eqnarray} \label{mech2B}
m_2\ir &\mapsto& m_2\uv=m_2\ir-1, \qquad \sigma=-1, \nonumber \\[2mm]
{I^{(1)}_k}\ir &\mapsto & {I^{(1)}_{k}}\uv={I^{(1)}_{k}}\ir-1,
\qquad k=1,\ldots,m_1, \\[2mm]
{I^{(2)}_k}\ir &\mapsto & {I^{(2)}_k}\uv={I^{(2)}_k}\ir+1,
\qquad k=1,\ldots,m_2\ir-1, \nonumber \\[2mm]
n_1\ir &\mapsto& n_1\uv=n_1\ir-1, \nonumber\\[2mm]
n_2\ir &\mapsto& n_2\uv=n_2\ir+1. \nonumber
\end{eqnarray}
\item[C. ${I^{(2)}_{m_2\irt}}\ir>0$:]
\begin{eqnarray} \label{mech2C}
m_2\ir &\mapsto& m_2\uv=m_2\ir+1, \qquad \sigma=-1, \qquad
{I^{(2)}_{m_2\uvt}}\uv=0, \nonumber \\[2mm]
{I^{(2)}_k}\ir &\mapsto & {I^{(2)}_k}\uv={I^{(2)}_k}\ir-1,
\qquad k=1,\ldots,m_2\ir, \\[2mm]
n_2\ir &\mapsto& n_2\uv=n_2\ir-1. \nonumber
\end{eqnarray}
\end{list}
where no label IR or UV is used for parameters that do not change during 
the flow.

An important consistency check is to show that the three mechanisms respect 
the counting of states at the endpoints of the flow.
From (\ref{mech2A}, \ref{mech2B}, \ref{mech2C}) one sees that for each
IR state there is precisely one applicable mechanism 
so that the IR counting of states is complete.
Moreover, using the recoursive identities among Gaussian polynomials
(\ref{ricors1}, \ref{ricors2})
the IR partition function in (\ref{partf}) naturally splits into six terms 
precisely associated with the three mechanisms, one for each sector:
\begin{eqnarray}
q^{\frac{c}{24}} Z^{\ir}_N(q) & =&  \sum_{(1,1) \text{A}}
q^{\frac{1}{4}{\boldsymbol{m}\irt}C\boldsymbol{m}\irt}
\gauss{m_1+n_1\ir-1}{m_1-1} \gauss{m_2\ir+n_2\ir-2}{m_2\ir-1} \nonumber
\\[1mm]
&&~ +\sum_{(1,1)\text{B}}
q^{\frac{1}{4}{\boldsymbol{m}\irt}C\boldsymbol{m}\irt}q^{m_1}
\gauss{m_1+n_1\ir-1}{m_1}  \gauss{m_2\ir+n_2\ir-2}{m_2\ir-1}\nonumber
\\[1mm]
&&~ + \sum_{(1,1)\text{C}}
q^{\frac{1}{4}{\boldsymbol{m}\irt}C\boldsymbol{m}\irt} q^{m_2\irt}
\gauss{m_1+n_1\ir}{m_1} \gauss{m_2\ir+n_2\ir-2}{m_2\ir} \label{irpart}
\\[1mm]
&&~ + \sum_{(3,1)\text{A}}
q^{\frac{1}{4}{\boldsymbol{m}\irt}C\boldsymbol{m}\irt-\frac{1}{2}} q^{n_2\irt}
\gauss{m_1+n_1\ir-1}{m_1-1} \gauss{m_2\ir+n_2\ir-2}{m_2\ir-2} \nonumber
\\[1mm]
&&~ + \sum_{(3,1)\text{B}}
q^{\frac{1}{4}{\boldsymbol{m}\irt}C\boldsymbol{m}\irt-\frac{1}{2}}
q^{n_2\irt+m_1}
\gauss{m_1+n_1\ir-1}{m_1}  \gauss{m_2\ir+n_2\ir-2}{m_2\ir-2}
\nonumber\\[1mm]
&&~ + \sum_{(3,1)\text{C}}
q^{\frac{1}{4}{\boldsymbol{m}\irt}C\boldsymbol{m}\irt-\frac{1}{2}} 
q^{n_2\irt+m_2\irt}
\gauss{m_1+n_1\ir}{m_1} \gauss{m_2\ir+n_2\ir-2}{m_2\ir-1}.
\nonumber
\end{eqnarray}
The labels $(1,1),~(3,1)$, A,~B,~C indicate that 
the corresponding sums on $m_1$, $m_2\irt$ are
restricted by the constraints imposed by the respective sector and  mechanism.
Notice that in the $(1,1)$ sector the quantum numbers in the second strip
are bounded by ${I_1^{(2)}}\ir \leqslant n_2\ir -1$ while in the sector 
$(3,1)$ the value ${I_1^{(2)}}\ir= n_2\ir$ is fixed 
and doesn't enter the combinatorics
that is equivalent to say that $m_2\ir-1$ zeros must be considered.
The two sum constrained by A can be understood using (\ref{Izero}) for both
the $q$-binomial factors and correspond to summing
under the constraint $I^{(1)}_{m_1}={I^{(2)}_{m_2\irt}}\ir=0$.

Similarly, the sum on B becomes apparent using (\ref{Ipos}) for the first strip
$q$-binomial and (\ref{Izero}) for the second strip $q$-binomial
and likewise the sum on C uses (\ref{Ipos}) for the second strip only.

Two steps are now involved: the first is the mapping of energies, the second 
is the counting of states.
The IR energy expression is given by (\ref{critenergy}). For the UV 
sector $(2,2)$ is given here \cite{OPW}:
\begin{equation} \label{crit22}
E=-\frac{c}{24}+\Delta_{2,2}+\frac{\boldsymbol{m}^{T}C\boldsymbol{m}}{4} 
-\frac{\sigma}{2}(m_1-m_2) +\sum_{i=1,2}\sum_{k=1}^{m_i} I_k^{(i)}, 
\end{equation}
$\Delta_{2,2}=\frac{3}{80}$; the rules for the various parameters are given
in Table~\ref{critTBA}.
An IR energy level at the base of a tower of states with string 
content fixed by
$(m_1,m_2\ir)$ maps to a UV energy level according to the energy
expressions (\ref{critenergy}, \ref{crit22}) that hold at the two 
conformal endpoints of the flow
\begin{equation} \label{energyABC}\hspace*{-3mm}
\begin{array}{ll} (1,1)  &\left\{ 
\begin{array}{l@{\hspace{7mm}}l@{~~\mapsto~~}l}
\mbox{A:} & q^{\frac{1}{4}\boldsymbol{m}\irt C\boldsymbol{m}\irt}
& q^{\Delta_{2,2}} \,
q^{\frac{1}{4}\boldsymbol{m}\uvt C\boldsymbol{m}\uvt} \,
q^{-\frac{1}{2}(m_1-m_2\uvt)}
\\[2mm]
\mbox{B:} & 
q^{\frac{1}{4}\boldsymbol{m}\irt C\boldsymbol{m}\irt} \, q^{m_1}
& q^{\Delta_{2,2}} \,
q^{\frac{1}{4}\boldsymbol{m}\uvt C\boldsymbol{m}\uvt} \,q^{m_2\uvt} 
q^{+\frac{1}{2}(m_1-m_2\uvt)} 
\\[2mm]
\mbox{C:} & 
q^{\frac{1}{4}\boldsymbol{m}\irt C\boldsymbol{m}\irt} \, q^{m_2\irt}
& q^{\Delta_{2,2}} \,
q^{\frac{1}{4}\boldsymbol{m}\uvt C\boldsymbol{m}\uvt} \,
q^{+\frac{1}{2}(m_1-m_2\uvt)}
\end{array} \right. \\[11mm] (3,1)  &\left\{
\begin{array}{l@{\hspace{7mm}}l@{~~\mapsto~~}l}
\mbox{A:} & 
q^{\frac{1}{4}\boldsymbol{m}\irt C\boldsymbol{m}\irt-\frac{1}{2}+n_2\irt}
& q^{\Delta_{2,2}} \,
q^{\frac{1}{4}\boldsymbol{m}\uvt C\boldsymbol{m}\uvt+n_2\uvt} \,
q^{-\frac{1}{2}(m_1-m_2\uvt)}
\\[2mm]
\mbox{B:} & 
q^{\frac{1}{4}\boldsymbol{m}\irt C\boldsymbol{m}\irt-\frac{1}{2}+n_2\irt} \, 
q^{m_1}
& q^{\Delta_{2,2}} \,
q^{\frac{1}{4}\boldsymbol{m}\uvt C\boldsymbol{m}\uvt+n_2\uvt-1} \,q^{m_2\uvt} 
q^{+\frac{1}{2}(m_1-m_2\uvt)} 
\\[2mm]
\mbox{C:} & 
q^{\frac{1}{4}\boldsymbol{m}\irt C\boldsymbol{m}\irt-\frac{1}{2}+n_2\irt} 
\, q^{m_2\irt}
& q^{\Delta_{2,2}} \,
q^{\frac{1}{4}\boldsymbol{m}\uvt C\boldsymbol{m}\uvt+n_2\uvt} \,
q^{+\frac{1}{2}(m_1-m_2\uvt)}.
\end{array} \right. \end{array}
\end{equation}
Using (\ref{mech2A}) to (\ref{mech2C}), the $q$-binomials
appearing in the IR partition function (\ref{irpart}) are rewritten in terms
of UV parameters. Also taking into
account the mapping of the energies (\ref{energyABC}) one obtains:
\begin{eqnarray} q^{\frac{c}{24}}Z_{N}^{\ir}(q)
& \mapsto & q^{\Delta_{2,2}} \left\{ \sum_{(1,1)\text{A}}
q^{\frac{1}{4}{\boldsymbol{m}\uvt}C\boldsymbol{m}\uvt} \,
q^{-\frac{1}{2}(m_1-m_2\uvt)}
\gauss{m_1+n_1\uv-1}{m_1-1}  \gauss{m_2\uv+n_2\uv-1}{m_2\uv}
\right. \nonumber \\
&& +  \sum_{(1,1)\text{B}}
q^{\frac{1}{4}{\boldsymbol{m}\uvt}C\boldsymbol{m}\uvt}\,
q^{\frac{1}{2}(m_1-m_2\uvt)} q^{m_2\uvt}
\gauss{m_1+n_1\uv}{m_1} \gauss{m_2\uv+n_2\uv-2}{m_2\uv}
\nonumber \\
&&+ \sum_{(1,1)\text{C}}
q^{\frac{1}{4}{\boldsymbol{m}\uvt}C\boldsymbol{m}\uvt} \,
q^{\frac{1}{2}(m_1-m_2\uvt)}
\gauss{m_1+n_1\uv}{m_1} \gauss{m_2\uv+n_2\uv-2}{m_2\uv-1} 
\label{intermed}\\
&& + \sum_{(3,1)\text{A}}
q^{\frac{1}{4}{\boldsymbol{m}\uvt}C\boldsymbol{m}\uvt} \,
q^{-\frac{1}{2}(m_1-m_2\uvt)}
\gauss{m_1+n_1\uv-1}{m_1-1}  \gauss{m_2\uv+n_2\uv-1}{m_2\uv-1}
\nonumber \\
&& + \sum_{(3,1)\text{B}}
q^{\frac{1}{4}{\boldsymbol{m}\uvt}C\boldsymbol{m}\uvt}\,
q^{\frac{1}{2}(m_1-m_2\uvt)} q^{n_2\uvt-1+m_2\uvt}
\gauss{m_1+n_1\uv}{m_1} \gauss{m_2\uv+n_2\uv-2}{m_2\uv-1}
\nonumber \\
&& + \left. \sum_{(3,1)\text{C}}
q^{\frac{1}{4}{\boldsymbol{m}\uvt}C\boldsymbol{m}\uvt} \,
q^{\frac{1}{2}(m_1-m_2\uvt)+n_2\uvt}
\gauss{m_1+n_1\uv}{m_1} \gauss{m_2\uv+n_2\uv-2}{m_2\uv-2} \right\}.
\nonumber 
\end{eqnarray}
The two terms corresponding to each mechanism can be summed with 
(\ref{ricors2}) so that the distinction among sectors disappear
\begin{eqnarray} q^{\frac{c}{24}}Z_{N}^{\ir}(q)
& \mapsto & q^{\Delta_{2,2}} \left\{ \sum_{\text{A}}
q^{\frac{1}{4}{\boldsymbol{m}\uvt}C\boldsymbol{m}\uvt} \,
q^{-\frac{1}{2}(m_1-m_2\uvt)}
\gauss{m_1+n_1\uv-1}{m_1-1}  \gauss{m_2\uv+n_2\uv}{m_2\uv}
\right. \nonumber \\
&& +  \sum_{\text{B}}
q^{\frac{1}{4}{\boldsymbol{m}\uvt}C\boldsymbol{m}\uvt}\,
q^{\frac{1}{2}(m_1-m_2\uvt)} q^{m_2\uvt}
\gauss{m_1+n_1\uv}{m_1} \gauss{m_2\uv+n_2\uv-1}{m_2\uv}
\nonumber \\
&& \left. + \sum_{\text{C}}
q^{\frac{1}{4}{\boldsymbol{m}\uvt}C\boldsymbol{m}\uvt} \,
q^{\frac{1}{2}(m_1-m_2\uvt)}
\gauss{m_1+n_1\uv}{m_1} \gauss{m_2\uv+n_2\uv-1}{m_2\uv-1} \right\} 
\nonumber \\
& =& 
 q^{\Delta_{2,2}} \sum_{\sigma=\pm 1}\sum_{m_1,m_2\uvt}
q^{\frac{1}{4}{\boldsymbol{m}\uvt}C\boldsymbol{m}\uvt} \,
q^{-\frac{\sigma}{2}(m_1-m_2\uvt)}
\gauss{m_1+n_1\uv-\delta_{\sigma,1}}{m_1-\delta_{\sigma,1}}
\gauss{m_2\uv+n_2\uv}{m_2\uv} \nonumber \\
&=&  q^{\frac{c}{24}} \chi_{2,2}^{(N)}(q)
\end{eqnarray}
and one is left with the UV partition function, that is the $(2,2)$
finitized character.
This argument shows that the IR counting of states with the three mechanisms 
induces the correct UV counting of states.

\section{The flow: TBA equations\label{s_TBA}}

The TBA equations are obtained with the procedure given in \cite{FPR2}: 
the double row 
transfer matrix (\ref{drtm}) eigenvalues can be normalized such that they 
satisfy a functional equation. The analytic content of the 
normalized transfer matrix eigenvalues is used to solve by Fourier transform 
the functional equation. As last step, a scaling limit is performed
to obtain the equations for a continuum theory (TBA equations).
The normalization factor is highly non-trivial in the 
sense that it contains both the spectral and the boundary parameter. 
The latter appears by a factor
$g(u,\xilatt)$ that is common to all the ``variable $r$'' cases and, 
with reference to the center lines of the two analyticity strips and at the
scaling limit, is given by:
\begin{eqnarray}
\label{hatg1v} \hat{g}_1(x,\xi) &= & \tanh \frac{x+\xi}{2}, \\
\label{hatg2v} \hat{g}_2(x,\xi) &= & 1.
\end{eqnarray}
This procedure leads to the same TBA equations, energy
expression and quantization conditions known \cite{FPR2} for 
the ``variable $r$'' flows:
\begin{eqnarray}
\label{tba1} \epsilon_1(x) &=& -\log \hat{g}_1(x,\xi)
-\sum_{k=1}^{m_1} \log (\tanh \frac{y_k^{(1)} -x}{2})
- K * L_2, \\
\label{tba2} \epsilon_2(x) &=& 4 e^{-x}
- \log \hat{g}_2(x,\xi) - \sum_{k=1}^{m_2} \log (\tanh \frac{y_k^{(2)} -x}{2})
-K * L_1
\end{eqnarray}
where the integration kernel is given by $K(x)=\frac{1}{2\pi \cosh x}$
and $L_j(x)=\log | 1+s_j \exp(-\epsilon_j)|$. The $*$ denotes the
convolution, $(f*g)(x)=\int_{-\infty}^{+\infty} dy\, f(x-y)g(y)$.
The energy expression is
\begin{equation}\label{scalingenergy}
E(\xi)=
\displaystyle \frac{2}{\pi} \sum ^{m_{1}}_{k=1}e^{-y^{(1)}_{k}}-
\int ^{\infty }_{-\infty }\! \! \frac{dx}{\pi^2}\,e^{-x} L_2
\end{equation}
and the quantization conditions are expressed in terms of the counting 
functions \cite{FPR2} $\psi_j(x)=-i\epsilon_j(x-i\frac{\pi}{2})$: 
\begin{eqnarray}\label{quant1}
\psi_2(y_k^{(1)})= -i\,\epsilon_2(y_k^{(1)}-i\frac{\pi}{2}) =
n_k^{(1)} \pi,  &~ & n_k^{(1)} = 2(I_k^{(1)}+m_1-k)+1-m_2,  \\[2mm]
\label{quant2}
\psi_1(y_k^{(2)})= -i\,\epsilon_1(y_k^{(2)}-i\frac{\pi}{2}) =
n_k^{(2)} \pi, & ~& n_k^{(2)} = 2(I_k^{(2)}+m_2-k)+1-m_1 .
\end{eqnarray}
Note the inversion of the indices: $\epsilon_1$ is for strip 2 and 
$\epsilon_2$ for strip 1.
This set of equations gives the complete description of the scaling energy 
for all the states of the flow under examination.
The correct choice of the parameters $m_i,~n_2$ is suggested by the lattice 
(Section~\ref{s_critpnt}) and a quick numerical check on the ground state 
leads to the values of the integration constants $s_i$. In summary, 
starting close to 
the IR point ($\xi\rightarrow +\infty $) and moving along the flow
one has:\\[3mm] \hspace*{3mm}
\begin{tabular}{l}
{\bf mechanism A and B, mechanism C before the collapse:}  \\[2mm]
\hspace*{7mm} 
$m_1, ~m_2 \mbox{ ~even}, \qquad n_2=\frac{m_1}{2}-m_2+1 \geqslant 0 ,\qquad
s_1=s_2=1, $ \\[3mm]
{\bf mechanism C after the collapse point:} \\[2mm]
\hspace*{7mm} 
$ m_2^A=m_2+2, \qquad n^{(2)}_{m_2^A-1}=n^{(2)}_{m_2^A}=1-m_1, $
\qquad the other variables remain unchanged.
\end{tabular} \\[4mm]
\begin{table}[t] \caption{\small\label{critTBA}
Classification of the patterns of 1- and 2-strings at the
endpoints of the flow. 
The parity $\sigma=\pm 1$ occurs in $(2,2)$ because there are frozen zeros.
The parities $s_1,s_2=\pm 1$ occur in the TBA equations as integration 
constants (see Appendix C). The 
expressions for $n_1$ are only used on a finite lattice because in
the scaling limit $n_1 \sim N/2 \rightarrow \infty$. The number of 
faces in a row, $N,$ is even.}\vspace*{-4mm}
$$
\begin{array}{|c|l|l|l|}
\hline
\chi_{r,s}(q)&\mbox{$(\boldsymbol{m},\boldsymbol{n})$ 
system}&\mbox{int. const.}&\mbox{quantum numbers}\\
\hline
\chi_{2,2}^{(N)}(q) &
\begin{array}{l}  m_{1} \mbox{~even},\, m_{2} \mbox{~odd} \\ 
n_2=(m_1-\sigma+1)/2-m_2 \\
n_1=(N+m_2+\sigma)/2-m_1 \end{array}&
\begin{array}{l}  s_{1}=-1 \\ s_{2}=1  \end{array} &
\begin{array}{l}
n^{(1)}_{k}=2(I_{k}^{(1)}+m_{1}-k)+1-m_{2}-\sigma  \\[2mm]
n^{(2)}_{k}=2(I_{k}^{(2)}+m_{2}-k)+1-m_{1}+\sigma
\end{array}\\
\hline
\begin{matrix} \chi_{1,1}^{(N)}(q) \\[2mm]+\chi_{3,1}^{(N)}(q) \end{matrix} &
\begin{array}{l} 
(1,1): ~~ m_{1}, ~m_{2}  \mbox{~even} \\
(3,1): ~~ m_1 ~\mbox{even},~m_2~\mbox{odd} \\
n_2=m_1/2-m_2+1 \\
n_1=(N+m_2)/2-m_1 \end{array} &
\begin{array}{l}
 s_{2}=1   \\ s_{2}=-1 \\[2mm]
 s_1=1 \rule[-4.1mm]{0mm}{3mm}
\end{array} 
&
\begin{array}{c}
n^{(1)}_{k}=2(I_{k}^{(1)}+m_{1}-k)+1-m_{2} \\[2mm]
n^{(2)}_{k}=2(I_{k}^{(2)}+m_{2}-k)+1-m_{1}
\end{array}\\ \hline
\end{array} $$
\end{table}
\noindent A small surprise is given at the IR point itself: performing the 
limit $\xi\rightarrow +\infty$ on the TBA equations leads to two different 
setups for the two sectors, namely the previous summary is  
correct for both the sectors along the flow and for the sector $(1,1)$ only
at the IR point. The sector $(3,1)$ at the IR point requires an odd number
of 1-strings in strip 2 as well as $s_2=-1$, as shown in Table~\ref{critTBA}.
This odd behaviour of the flow close to the IR point is new:
in all the previous cases, the movement off the IR point was observed to be 
smooth and the content of zeros never came up with a change in the value 
of $m_2$.
The reason can be easily found in the discontinuity of the following limit
(eq. (3.70) of \cite{FPR2}): 
\begin{equation} \label{limits}
\lim_{x\rightarrow-\infty} \psi_1(x)=
\lim_{x\rightarrow-\infty} i\log \hat{g}_1(x,\xi)=
\left\{ \begin{array}{l@{\mbox{~~if~~}}l} \pi &  \xi<+\infty, \\
0 & \xi=+\infty \end{array} \right.
\end{equation}
In the first case, the odd parity of $n_k^{(2)}$ (\ref{quant2}) implies that
a second strip zero can be found at $-\infty$. Indeed, a clear numerical 
evidence and the following simple argument show that $y_1^{(2)}=-\infty$ 
precisely for all the states in the $(3,1)$ sector.
Looking at the difference between the endpoints of $\psi_1$ 
along the flow (assuming $y_{m_1}^{(1)}<+\infty$) one has
\begin{equation}
\lim_{x\rightarrow-\infty} \psi_1(x)-\lim_{x\rightarrow+\infty} \psi_1(x)=
\pi (1+ m_1) = \pi \cdot \mbox{``odd''}
\end{equation}
so that the maximum number of zeros (1- and 2-strings) that find room in that 
interval\footnote{$\lfloor x\rfloor$ means the integer part of $x$.} is 
$\lfloor \frac{m_1+1}{2}\rfloor+1=\frac{m_1}{2}+1$ that 
coincides with the actual number of 1- and 2-strings in the strip, $m_2+n_2$.
So, all the locations are full and when $I_1^{(2)}=n_2$, equivalent
to the $(3,1)$ sector, the lowest location must be occupied by a 1-string.
Observe that the alternative assumption 
$y_{m_1}^{(1)}=+\infty$ leads to the same conclusion because $m_1$ is even. 
Summarizing, along the flow one has $y_1^{(2)}=-\infty$; in the IR limit 
$\xi \rightarrow +\infty$ this zero must leave the complex plane 
because of the second case in (\ref{limits}) and this leads to the rules for 
the TBA in $(3,1)$ sector, as in Table~\ref{critTBA}.

\begin{figure}\begin{center}
\includegraphics[width=0.85\linewidth ]{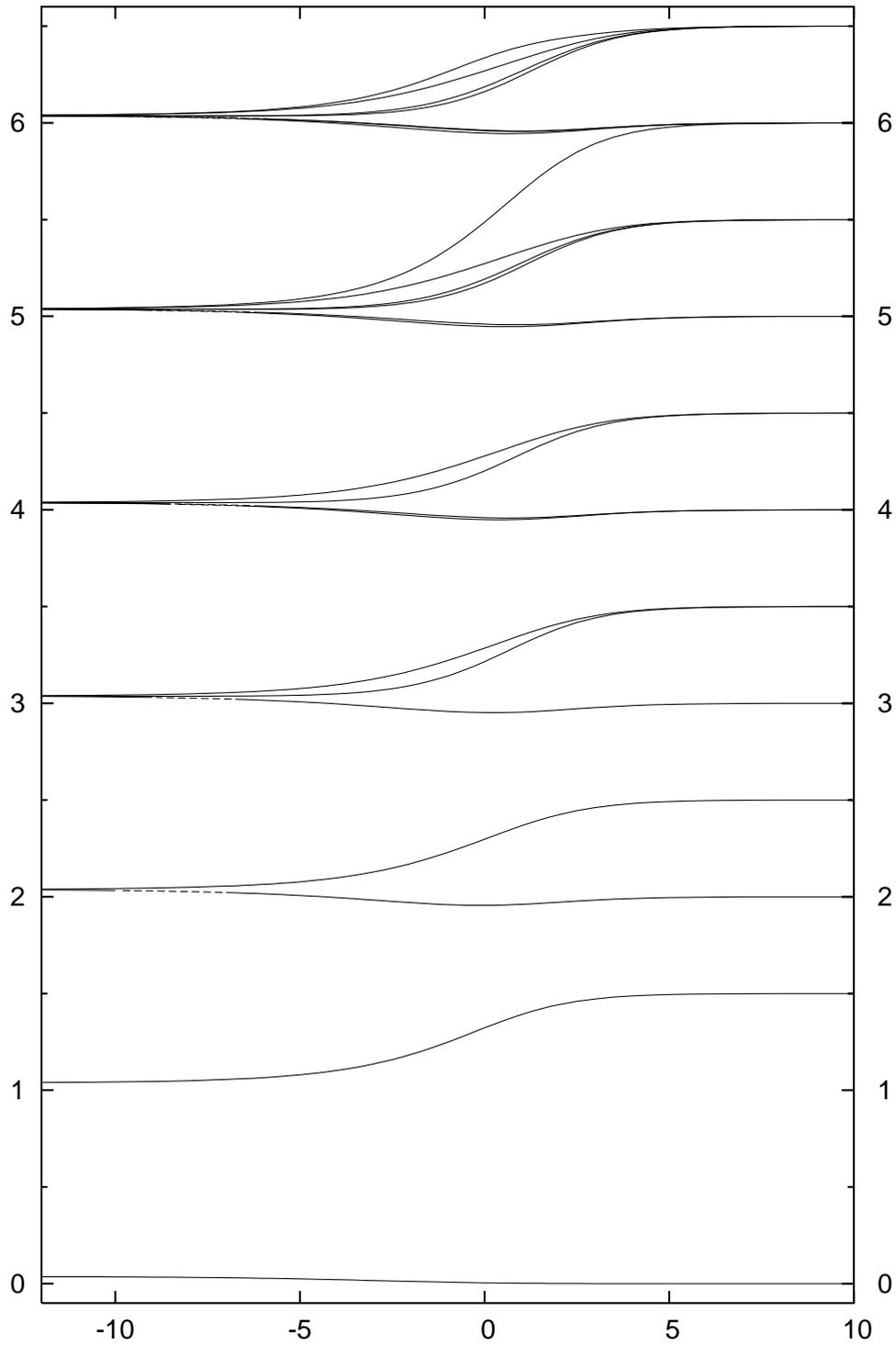}
\caption{\small Scaling energies for the flow $\chi_{2,2} \mapsto \chi_{1,1}+
\chi_{3,1}$.
The list of states is given in Table~\ref{t22_1131}. The intermediate region
of the mechanism~C levels (shown dashed) are schematic and have not
been obtained from the solution of the TBA equations.
\label{energyflow}}
\end{center}
\end{figure}

A numerical solution of this set of TBA equations for the first few 
states was done and is given in Fig.~\ref{energyflow}.

\section{Conclusions}
The lattice approach to boundary renormalization flows exposed in 
\cite{FPR2} and extended here appears quite effective in 
obtaining the scaling energy for all the excitations.
It should be interesting to understand something more about the role played 
by the picture of zeros in the continuum theory itself. 
Indeed, this structure survives the scaling limit and 
is relevant in the description of states. 

Moreover, there are structures that are common to 
the $\varphi_{1,3}$ perturbations \cite{PCA} of tricritical Ising model. 
They are the types of zeros and number of strips, 
corresponding to the number of pseudoenergies,
the way these pseudoenergies are coupled to form the TBA system and
the kernel.
It should be nice to have track of this from the 
continuum theory itself.

It is also an interesting open question to extend the present
lattice approach to all the boundary flows for the whole 
series A-D-E of minimal models.

\section*{Acknowledgments}
The author would like to thank P. Pearce and F. Ravanini for 
having read the manuscript.
The content of Appendix C comes from a private discussion with P. Pearce.

\appendix
\renewcommand{\theequation}{A.\arabic{equation}}
\setcounter{equation}{0}

\section*{Appendix A. Gaussian polynomials}
The finitized characters are know to be expressed in terms of 
$q$-binomials. They are combinatorial expressions defined by:
\begin{eqnarray}\label{gauss}
\gauss{m+n}{m} &=&
\sum_{I_1=0}^n \sum_{I_2=0}^{I_1}\cdots \sum_{I_m=0}^{I_{m-1}}
q^{I_1+\ldots+I_m}. 
\end{eqnarray}
Two recoursive relations are important for this paper:
\begin{eqnarray}\label{ricors1}
\gauss{m+n}{m} &=&\gauss{m+n-1}{m-1} + q^m \gauss{m+n-1}{m},\\[4mm]
\label{ricors2}
\gauss{m+n}{m} &=&\gauss{m+n-1}{m} + q^n \gauss{m+n-1}{m-1}.
\end{eqnarray}
The following expressions are useful to understand the contribution of
each mechanism to the full character. Indeed, the restriction $I_m>0$
leads to
\begin{equation} \label{Ipos}
\sum_{I_1=1}^{n} \sum_{I_2=1}^{I_1}
\ldots \sum_{I_{m}=1}^{I_{m-1}} q^{I_1+\ldots+I_{m}}  =
q^{m} \sum_{I'_1=0}^{n-1} \sum_{I'_2=0}^{I'_1}
\ldots \sum_{I'_{m}=0}^{I'_{m-1}}
q^{I'_1+\ldots+I'_m} =  q^{m} \gauss{m+n-1}{m}
\end{equation}
whereas the restriction $I_m=0$ leads to
\begin{equation} \label{Izero}
\sum_{I_1=0}^{n} \sum_{I_2=0}^{I_1}
\ldots \sum_{I_{m-1}=0}^{I_{m-2}} q^{I_1+\ldots+I_{m-1}}
= \gauss{m+n-1}{m-1}.
\end{equation}

\renewcommand{\theequation}{B.\arabic{equation}}
\setcounter{equation}{0}

\section*{Appendix B. Proof of the expression for the energy at 
$(1,1)\oplus(3,1)$}
The energy expression (\ref{critenergy}) is proved here 
using the continuum energy and the TBA equations in the limit 
$\xi\rightarrow +\infty$. 
The method is similar to the computation done in \cite{OPW} 
at each critical point but, in the present case, 
one has to consider extra terms that appear because the limit 
$\xi\rightarrow +\infty$ will be taken only at the end of the computation.

The first step is to transform (\ref{scalingenergy})
in a more explicit form. Its first term, $e^{-y_k^{(1)}}$,
can be expressed by the quantization condition (\ref{quant1}) and 
then one adds  
$0=\sum_{k=1}^{m_2} (n_k^{(2)}\pi-\psi_1(y_k^{(2)}))$ from (\ref{quant2}).  
Some numeric terms are obtained, given by sums and products of $m_i$ and 
integer numbers. They can be easily worked out with the help of the 
Cartan matrix defined in (\ref{critenergy}) leading to
\begin{eqnarray} \label{enel}
E(\xi) &=&\frac{\boldsymbol{m}^{T}C\boldsymbol{m}}{4}+\sum_{i=1}^{2}
\sum_{k=1}^{m_i} I_k^{(i)} 
-\frac{i}{2\pi}\sum_{k=1}^{m_2}\log \hat{g}_1(y_k^{(2)}-i\frac{\pi}{2},\xi)\\
&&+ \int_{-\infty}^{+\infty} \frac{dx}{4\pi^2}\left[\sum_{k=1}^{m_1}
\frac{L_1(x)}{\sinh(y_k^{(1)}-x)}+\sum_{k=1}^{m_2}\frac{L_2(x)}
{\sinh(y_k^{(2)}-x)}-4 e^{-x} L_2(x) \right]. \nonumber
\end{eqnarray}
The contribution from the integration can be worked out with the 
following procedure.
The main point is to compute, in two different ways, the real integral:
\begin{eqnarray} \label{reals}
S_j&=&\int_{-\infty}^{+\infty}dx \left[ -\epsilon_j' L_j+\re(\epsilon_j)L_j'
\right] \\
&=& \int_{-\infty}^{+\infty}dx \left[ (\log s_j \hat{t}_j)' \log|1+\hat{t}_j|
-\re(\log (s_j \hat{t}_j)) (\log|1+\hat{t}_j|)' \right] \nonumber
\end{eqnarray}
where $\hat{t}_j=s_j\exp(-\epsilon_j)$.
On the first hand, the real axis is divided in the following intervals
\begin{equation} \label{interv}
\int_{-\infty}^{+\infty}=\int_{-\infty}^{y_1^{(j)}}
+\int_{y_1^{(j)}}^{y_2^{(j)}}+\ldots \int_{y_k^{(j)}}^{\xi}+
\int_{\xi}^{y_{k+1}^{(j)}}+\ldots 
+\int_{y_{m_j}^{(j)}}^{+\infty}
\end{equation}
(in strip 2 there is no need to consider intervals with $\xi$). 
The integration variable can be changed to
be $t=\hat{t}_j$, with $dt=\hat{t}_j' dx$ and in this new variable it is 
apparent that
all the contributions corresponding to the intervals (\ref{interv})
vanish because $\hat{t}_j(y_k^{(j)})=0$, except the first and the last. 
This leads to
\begin{equation}
S_j= \int_{\hat{t}_j(-\infty)}^{\hat{t}_j(+\infty)}dt 
\left( \frac{\log |1+t|}{t} -\frac{\log |t|}{1+t} \right). 
\end{equation}
Along the flow ($\xi \in \mathbb{R}$), the required endpoints of the integral 
take the following values: 
\begin{equation}\label{limiti}
\begin{array}{l}
\displaystyle \hat{t}_j(+\infty)=2\cos \lambda=\frac{1+\sqrt{5}}{2},  \\[2mm]
\hat{t}_1(-\infty)=s_1\log \hat{g}_1(-\infty,\xi)=-1,\qquad 
\hat{t}_2(-\infty)=0.
\end{array}\end{equation}
The integral itself is recognised to be a sum of Rogers dilogaritms
that, at the previous endpoints, can be exactly computed. Actually
only the sum is required: 
\begin{equation} \label{somma}
S_1+S_2=\frac{11}{15}\pi^2.
\end{equation}
On the other hand, one can substitute in (\ref{reals}) the expressions 
for $\epsilon_j$, $\epsilon_j'$ given by the TBA equations (\ref{tba1}, 
\ref{tba2}). In doing this, all terms involving the 
convolution have cancelled because of the symmetry of the kernel, 
$K(x)=K(-x)$. 
Among the remaining terms, those involving the derivatives $L_j'$ are 
integrated by parts leading to
\begin{equation}
S_1+S_2 = -2\int_{-\infty}^{+\infty}dx \left[ 
\sum_{k=1}^{m_1} \frac{L_1}{\sinh(y_k^{(1)}-x)} 
+\sum_{k=1}^{m_2} \frac{L_2}{\sinh(y_k^{(2)}-x)} -4e^{-x} L_2\right] 
+2\int_{-\infty}^{+\infty}dx \frac{\hat{g}_1'}{\hat{g}_1}L_1 .
\end{equation}
The first term on the right hand side enters the expression for the 
energy (\ref{enel}).
Using the result (\ref{somma}) one obtain, for all $\xi \in \mathbb{R}$:
\begin{equation} \label{energia}
E(\xi) =-\frac{11}{120}+\frac{\boldsymbol{m}^{T}C\boldsymbol{m}}{4}+
\sum_{i=1}^{2}\sum_{k=1}^{m_j} I_k^{(j)} 
-\frac{i}{2\pi}\sum_{k=1}^{m_2}\log \hat{g}_1(y_k^{(2)}-i\frac{\pi}{2},\xi)
+ \int_{-\infty}^{+\infty} \frac{dx}{4\pi^2}
\frac{\hat{g}_1'}{\hat{g}_1}L_1. 
\end{equation}
This is the espression for the energy with the most explicit dependence on the 
boundary term $\hat{g}_1$ so that is well suited to the purpose of taking the
limit $\xi \rightarrow +\infty$. 
Using the explicit expression for $\hat{g}_1$ 
and remembering that, along the flow, $y_1^{(2)}=-\infty$ if and only if 
$I_1^{(2)}=n_2$, one obtains
\begin{equation}\label{special}
\displaystyle \lim_{\xi \rightarrow +\infty} 
-\frac{i}{2\pi}\sum_{k=1}^{m_2}\log \hat{g}_1(y_k^{(2)}-i\frac{\pi}{2},\xi)
=\lim_{\xi \rightarrow +\infty} -\frac{i}{2\pi}\sum_{k=1}^{m_2}\log \tanh 
(\frac{y_k^{(2)}+\xi}{2}-i\frac{\pi}{4}) 
=-\frac{1}{2} \delta_{I_1^{(2)},n_2}. 
\end{equation}
This term is the new feature of the flow under examination
with respect to the cases with single character. 
In spite of the fact that  $\lim _{\xi \rightarrow +\infty} \hat{g}_1'=0$, 
the term under the integral also gives a contribution
because of the divergence $L_1(-\infty)=\log 0 =-\infty$ (\ref{limiti}). 
From the TBA equations the following asymptotic behaviour 
is obtained, for all fixed values of $\xi$
\begin{equation} \label{asint}
\epsilon_1(x) ~\mathop{\sim}_{x\sim-\infty} ~
-\log \tanh \frac{x+\xi}{2} .
\end{equation}
To get this from (\ref{tba1}), one considers that $y_k^{(1)} \gg -\infty$ 
because of the infinite number of 2-strings appearing in the scaling
limit \cite{FPR2}. In addition to that, $L_2(-\infty)=0$ was used. 
Actually, the following term will be used: 
\begin{equation}\label{asintL}
L_1'=-\frac{s_1 e^{-\epsilon_1}\epsilon_1'}{1+s_1 e^{-\epsilon_1}} 
{~\mathop{\sim}_{x\sim -\infty}~}
\frac{1}{2} (1+\tanh\frac{x+\xi}{2}).
\end{equation}
This asymptotic behaviour in the variable $x$ is relevant  
in the integral term in (\ref{energia}), when $\xi \rightarrow +\infty$.
This is apparent from the following explicit expression:
\begin{equation}
\mathcal{I} \equiv \int_{-\infty}^{+\infty} \frac{dx}{4\pi^2}
\frac{\hat{g}_1'}{\hat{g}_1}L_1 = 
-\int_{-\infty}^{+\infty} \frac{dx}{4\pi^2}
\log |\hat{g}_1| L_1' = -\int_{-\infty}^{+\infty} \frac{dx}{4\pi^2}
\log |\tanh \frac{x+\xi}{2} | L_1'
\end{equation}
where an integration by parts has been performed and the corresponding
boundary terms vanish.
One has: 
\begin{eqnarray}
\mathcal{I} &\displaystyle \mathop{\sim}_{\xi \sim +\infty} &
-\frac{1}{2}\int_{-\infty}^{+\infty} \frac{dx}{4\pi^2}
\log |\tanh \frac{x+\xi}{2}| (1+\tanh\frac{x+\xi}{2}) \nonumber \\
&=& -\int_{-\infty}^{+\infty} \frac{dx}{4\pi^2}
\log |\tanh x| (1+\tanh x) \nonumber \\
&=&
-\int_{-\infty}^{+\infty} \frac{dx}{4\pi^2}
\log |\tanh x| = -\int_{0}^{+\infty} \frac{dx}{2\pi^2}
\log \tanh x
\end{eqnarray}
where on the first line the asymptotic behaviour was used, 
in the second line a redefinition of $x$ shows that the integral
is independent on $\xi$, on the third line a term vanish because it
is odd under the exchange $x\rightarrow -x$; 
the remaining integrand is even under the same exchange 
and this leads to the last equality.
With a new change of variables, this definite integral 
is transformed in 
\begin{equation}\label{inte}
\mathcal{I}= -\int_{0}^{1} \frac{dt}{4\pi^2}
\frac{1}{t} \log \frac{1-t}{1+t}=\frac{1}{16}.
\end{equation}
The actual value can be explicitly computed or
obtained from standard mathematical tables of integrals;
it admits also an expression in terms of Rogers dilogarithms.
In conclusion, from (\ref{energia}, \ref{special}, \ref{inte})
one has the energy at the IR critical point $(1,1)\oplus(3,1)$
\begin{equation}
E =-\frac{7}{240}+\frac{\boldsymbol{m}^{T}C\boldsymbol{m}}{4}
-\frac{1}{2} \delta_{I_1^{(2)},n_2}+\sum_{i=1}^{2}\sum_{k=1}^{m_i} I_k^{(i)} 
\end{equation}
that is precisely the expression (\ref{critenergy}) given in
Section~\ref{s_critpnt}. The first term is the central charge contribution
$-\frac{c}{24}$. 

As a final comment, the expression (\ref{energia}) can be generalized 
to all the flows with the obvious addition of terms depending upon 
$\hat{g}_2$ and a possibly non-zero term that takes into account the 
subtleties related to $\hat{t}_1(-\infty)$.

\section*{Appendix C. Rule for the integration constants $s_j$.} 
\renewcommand{\theequation}{C.\arabic{equation}}
\setcounter{equation}{0}

The braid limit used in the first line of (\ref{limiti}) can be used to obtain 
the general selection rule for the variables $s_j$.
They appear in the derivation of the TBA equations as integration 
constants (see \cite{FPR2} for details).
Then, taking the limit $x\rightarrow +\infty$ in the equations themselves, 
one can use the braid limit to express the asymptotic values of 
$\epsilon_j,~L_j$. This leads to:
\begin{equation}
s_j=(-1)^{m_j}, \qquad \xi\in \mathbb{R}, ~\xi=+\infty=\text{IR}.
\end{equation}
An implicit assumption was used, that all the 
1-strings are in a finite position, $y_k^{(j)} <+\infty$, condition that is 
always satisfied except at the UV point. If this is violated, the 
actual value of the braid limit can change and the previous argument 
doesn't hold. 
The correct UV rule can be obtained by the previous one 
observing that if along the flow one has $\hat{g}_j\neq 1$, the content of 
zeros changes in the strip $3-j$ so that:
\begin{equation}
s_j=(-1)^{m_j+1}, \qquad \xi=-\infty=\text{UV}.
\end{equation}

The selection rules proven here hold for all the flows
and all the conformal boundary conditions.
Of course, they contain the cases given in \cite{FPR2}  
and, in this sense, this Appendix completes the exposition
given there.


\end{document}